# The Mathematical Modelling of Inhomogeneities in Ventricular Tissue

T.K. SHAJAHAN[1], SITABHRA SINHA[2], and RAHUL PANDIT[1,*]
[1]Centre for Condensed Matter Theory, Department of Physics,
Indian Institute of Science, Bangalore 560012, India.
[2] Institute of Mathematical Sciences, C I T Campus, Taramani, Chennai 600113, India

**Abstract**
Cardiac arrhythmias such as ventricular tachycardia (VT) or ventricular fibrillation (VF) are the leading cause of death in the industrialised world. There is a growing consensus that these arrhythmias arise because of the formation of spiral waves of electrical activation in cardiac tissue; unbroken spiral waves are associated with VT and broken ones with VF. Several experimental studies have been carried out to determine the effects of inhomogeneities in cardiac tissue on such arrhythmias. We give a brief overview of such experiments, and then an introduction to partial-differential-equation models for ventricular tissue. We show how different types of inhomogeneities can be included in such models, and then discuss various numerical studies, including our own, of the effects of these inhomogeneities on spiral-wave dynamics. The most remarkable qualitative conclusion of our studies is that the spiral-wave dynamics in such systems depends very sensitively on the positions of these inhomogeneities.

## 1 Introduction

Mammalian hearts are electromechanical pumps in which the propagation of electrical activation waves through cardiac tissue triggers the mechanical contraction of the walls of the atria or ventricles. The walls of these chambers are made of tissue, which is an excitable medium that supports the passage of regular contraction waves across it. This excitability depends crucially on the states of cardiac cells that make up the tissue. In its quiescent state, such a cell maintains a potential difference of ~ -85 mV across the cell membrane. If the cell is stimulated beyond a certain threshold potential (~ -60 mV), voltage-gated ion channels on the membrane allow $Na^+$ ions to enter the cell. Thus the cell is rapidly depolarized; it repolarizes partially for ~ 10 ms, and then enters a plateau state in which ion channels, for $Ca^{++}$ and $K^+$ ions, are activated. This leads to the repolarization of the cell and it returns eventually to its quiescent state. The repolarization phase lasts much longer than the initial

---

[*] Also at Jawaharlal Nehru Centre for Advanced Scientific Research, Bangalore 560 064, India

depolarization phase (hundreds of ms compared to a few ms). The depolarization-repolarization response is referred to as the action potential. The action potential duration (APD) in a human ventricular cell is typically ~ 200 ms; a typical plot of an action potential is shown in Fig. 1. During the period of repolarization and slightly after it, a cardiac cell cannot be easily excited again and is said to be in a refractory state. The coupling between cardiac cells results in the propagation of excitation from one cell to the other.

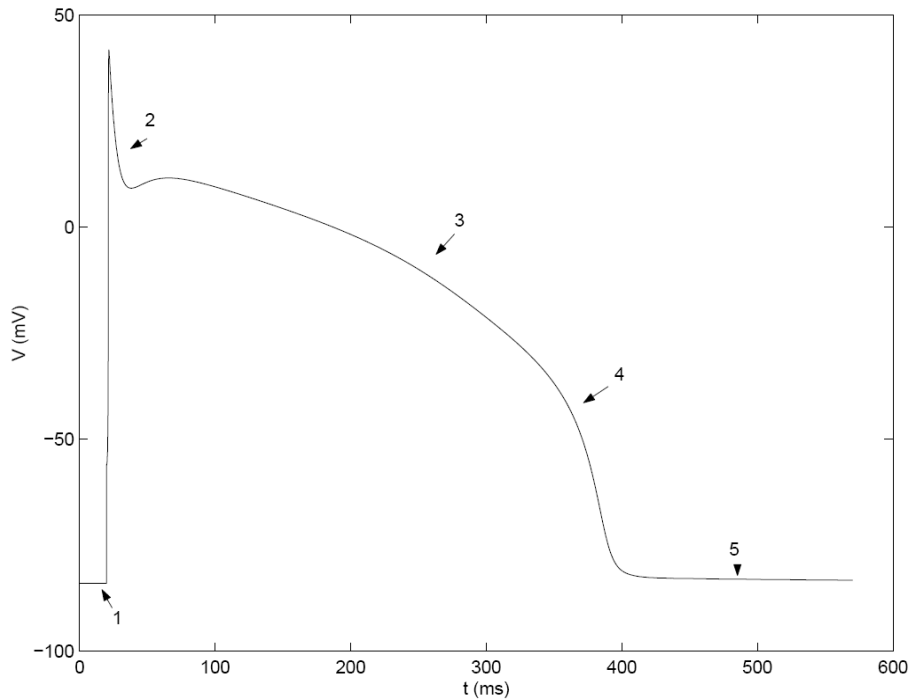

FIG. 1. An example of a cardiac action potential obtained from a single cell in the Luo-Rudy model showing (1) the action potential upstroke or rapid depolarization, (2) the rapid repolarization, (3) the plateau stage, (4) the final stage of repolarization, and (5) the resting state, in a plot of transmembrane potential $V$ versus time $t$.

In a normal heart, the regular rhythmic contraction of the atria and ventricles pumps blood through the body. Cardiac arrhythmias disturb this regular rhythm; the most dangerous of these are ventricular tachycardia (VT) and ventricular fibrillation (VF). During an episode of VT the rate at which a patient's heart beats increases to more than 100 beats per minute; an electrocardiogram (ECG) shows a roughly periodic trace but one that is considerably faster than the normal sinus rhythm; if VF occurs then, not only does the rate of heart beats increase, but the ECG trace seems quite chaotic. (An introduction to ECGs that is accessible to nonspecialists can be found at Ref. [1]) There is growing consensus that VT is associated with the formation of a single spiral wave of electrical activation in ventricular tissue whereas VF is believed to arise when such spiral waves break leading to spiral turbulence. Voltage-sensitive dyes and CCD cameras can be used to record the propagation of the such spiral waves of excitation in cardiac tissue[2-5]. Indeed spiral waves are ubiquitous in excitable media, in which subthreshold excitations decay rapidly but suprathreshold excitations lead to a response like the action potential described above. Cardiac tissue is one example of an excitable medium. Such excitable media are modeled by coupled, nonlinear, partial differential equations that describe the spatiotemporal evolution of concentration fields; in suitable parameter ranges these models can show regular spiral waves or spiral turbulence that is accompanied by spatiotemporal chaos [6]. In the next Section we will discuss two such models for cardiac tissue.

There are several mechanisms for the onset of spiral waves in excitable media. One mechanism relies on electrophysiological inhomogeneities, set up in the tissue because of the passage of the spiral wave; i.e., the refractory tissue in the wake of an activation wave can act as a temporary obstacle in the medium. Such reentry is referred as functional reentry. Another mechanism invokes the presence of permanent inhomogeneities in cardiac tissue, e.g., inexcitable cells, such as scar tissue caused by a previous cardiac arrest or ischaemia. Excitation waves can move around such permanent inhomogeneities; the resulting reentry is referred as anatomical reentry and it can lead to a spiral wave that moves around the obstacle.

Tissue heterogeneities, which can occur at several length scales (from microns to mm), can be of various types. A small percentage of cardiac tissue is fibrotic and non-excitable; this percentage can increase significantly (up to 30%-40%) during aging, after a myocardial infarction, or in cardiac myopathies. A myocardial infarction can leave slightly bigger obstacles of poorly excitable regions (of the order of mm). Major arteries in the heart can also act as obstacles in the path of a propagating wave. We concentrate on inhomogeneities that are large enough that partial-differential-equation models for cardiac tissue are valid.

Clearly the effects of inhomogeneities, functional and anatomical, on wave propagation and spiral-wave formation and break up in cardiac tissue is an important question that has a direct bearing on life-threatening cardiac arrhythmias. Not surprisingly then various experimental and theoretical studies have been undertaken to address this question. We begin with a very brief overview of studies that have concentrated on understanding the dynamics of spiral waves in two-dimensional cardiac tissue in which some inhomogeneities are introduced systematically.

Experiments conducted on ventricular tissue with a single obstacle have shown that small obstacles do not affect spiral waves but, as the obstacle size increases, a spiral wave can get attached to the obstacle and then start moving around it[7]. Davidenko *et al* [8] have found that, in one case, an artificially induced spiral wave moved away from its site of origin in their cardiac-tissue preparation because of an obstacle. By contrast, other studies[9-12] have shown that an obstacle, in the path of a moving spiral wave, can break it and lead to many competing spiral waves. Recent experiments by Hwang *et al*[13] have suggested that multistability of spirals with different periods in the same cardiac-tissue preparation can arise because of the interaction of spiral tips with small-scale inhomogeneities.

It has also been found in numerical simulations of two-dimensional models for cardiac tissue that such obstacles can convert a meandering (detached) spiral wave into a stable (attached) one [14-16]. A naturally occurring obstacle like papillary muscle can also play the role of an attaching structure during VT[17].

A systematic numerical study has been carried out for a circular obstacle (by creating a hole in the simulation domain for the Luo-Rudy model). This study shows that, if the hole is large enough, a spiral wave can get anchored to it; but, as the radius of the hole is decreased, there is a transition from periodic to quasiperiodic motion of spiral wave via a Hopf bifurcation, and eventually a transition to spiral-wave breakup and spatiotemporal chaos[14].

Starobin, *et al* [18] have studied, numerically and analytically, the interaction of spiral waves with piece-wise linear obstacles. They find that if the excitability of the medium is high, the wave moves around the obstacle boundary, rejoins itself and then proceeds as if it had not encountered any obstacles in its path. However, if the excitability is low, the two ends of this wavefront are unable to join, so two free ends survive, curl up and develop in to two spiral waves, which in turn can break up again. This study also finds that, apart from the excitability of the medium and the local curvature of the wave front, the shape of the obstacle also affects the attachment spiral to the obstacle.

Anchoring helps in the control of such spiral waves. In the anchored condition the spiral wave is considerably less dangerous than a broken spiral wave for the former is associated with VT whereas the latter is associated with VF. Furthermore, another numerical study has shown that [19] a single anchored spiral wave can be easily unpinned and removed from the simulation domain by applying a weak, uniform electric field (less than 0.5 V/cm). Such a method is very useful because one does not need to know the exact location of the spiral tip to eliminate it from the domain.

A numerical study of a FitzHugh-Nagumo-type model has shown that spiral break up can be suppressed by introducing a large fraction of non-excitable cells, distributed randomly in the simulation domain [20,21]. In the following Sections we give an overview of our numerical studies of spiral-wave dynamics in the presence of obstacles in two models of ventricular tissue. Our goal has been to design studies to bring out the dependence of such dynamics on the position, size, and shapes of such obstacles. Some details of this work have appeared in Ref. [22], where we have shown that spiral-wave dynamics depend sensitively on all these parameters; we also present some new results here for the interactions of spiral waves with two inhomogeneities. Section 2 is devoted to a description of the models and our numerical schemes. Section 3 contains an overview of our results. Section 4 concludes this paper with a discussion of the implications of our studies for cardiac arrhythmias and their control.

## Models

In this Section we give a brief description of the two mathematical models for ventricular tissue that we use in our studies. These are the Panfilov model [23,24] and the Luo-Rudy I (LRI) model [25]. The former belongs to simplified models of the FitzHugh-Naguma type; the latter is a detailed physiological models based on the Hodgkin-Huxley formalism. Such detailed models account for the ionic currents that are transported through voltage-gated ion channels in cardiac tissue, so their numerical simulation is computationally expensive. Simulations of simplified models like that of Panfilov [23,24], are not as demanding from the point of view of computational resources; however, they often suffice to obtain results that are qualitatively similar to those that follow from studies of detailed models like LRI[14, 26-29]. As we will see below, the ways in which spiral waves appear, break up, and interact with obstacles, and the way in which spiral turbulence can be controlled in the Panfilov model are qualitatively similar to those in the LRI model. Our discussion is based predominantly on recent studies [22, 26-29] that we have carried out in our group; our earlier studies of the control of spiral turbulence in these models can be found in Refs. [6, 29].

The Panfilov model consists of a partial differential equation (PDE) for the transmembrane potential $V(x,t)$ coupled to an ordinary differential equation for the recovery variable $g(x,t)$ at the point $x$ and time $t$. All information about the ion channels is lumped into $g$, which is related to the membrane conductance. The Panfilov-model equations are

$$\partial V / \partial t = \nabla^2 V - f(V) - g,$$
$$\partial g / \partial t = \varepsilon(V,g)(kV - g). \qquad (1)$$

The initiation of action potential arises because of the following piecewise-linear form for $f(V)$: $f(V) = C_1 V$, for $V < e_1$, $f(V) = -C_2 V + a$, for $e_1 \leq V \leq e_2$, and $f(V) = C_3 (V - 1)$, for $V > e_2$. We use physically appropriate parameters given in Refs. [23,24] are $e_1 = 0.0026$, $e_2 = 0.837$, $C_1 = 20$, $C_2 = 3$, $C_3 = 15$, $a = 0.06$, and $k = 3$. The function $\varepsilon(V,g)$ determines the dynamics of the recovery variable: $\varepsilon(V,g) = \varepsilon_1$ for $V < e_2$, $\varepsilon(V,g) = \varepsilon_2$ for $V > e_2$, and $\varepsilon(V,g) = \varepsilon_3$ for $V < e_1$ and $g < g_1$, with $g_1 = 1.8$, $\varepsilon_1 = 1/75$, $\varepsilon_2 = 1.0$, and $\varepsilon_3 = 0.3$.

We will also show representative results from our simulations of the Luo-Rudy I (LRI) model, which is based on the Hodgkin-Huxley formalism. It accounts for 6 ionic currents [25] (e.g., $Na^+$, $K^+$, and $Ca^{2+}$) and 9 gate variables for the voltage-gated ion channels that control ion movement across the cardiac cell membrane. In the resting state there is a potential difference of $\approx -84$ mV across this cell membrane. If a stimulus raises this above -60 mV, $Na^+$ channels open, resulting in an action potential (Fig.1). Cells in the LRI model are coupled diffusively; thus one must solve a PDE (see Appendix) for the transmembrane potential $V$; the time evolution and $V$ dependence of the currents that appear in this PDE are given by 7 coupled ordinary differential equations (ODEs) [22,25].

We restrict ourselves to two-dimensional studies here and refer the reader to Ref. [22] for our three-dimensional simulations. For both the Panfilov-model and LRI PDEs we use a forward-Euler method in time $t$ and a finite-difference method in space that uses a five-point stencils for the Laplacian. Our spatial grid consists of a square lattice with side $L$ mm. For the Panfilov model we use a temporal and spatial steps $\delta t = 0.022$ and $\delta x = 0.5$, respectively, and $L = 200$ mm. We take the dimensioned time [23,24] $T$ to be 5 ms times dimensionless time and 1 spatial unit to be 1 mm; the dimensioned value of the conductivity constant $D$ is 2 cm$^2$/s. For the LRI model we use $\delta t = 0.01$ ms, $\delta x = 0.0225$ cm, and $L=90$ mm.

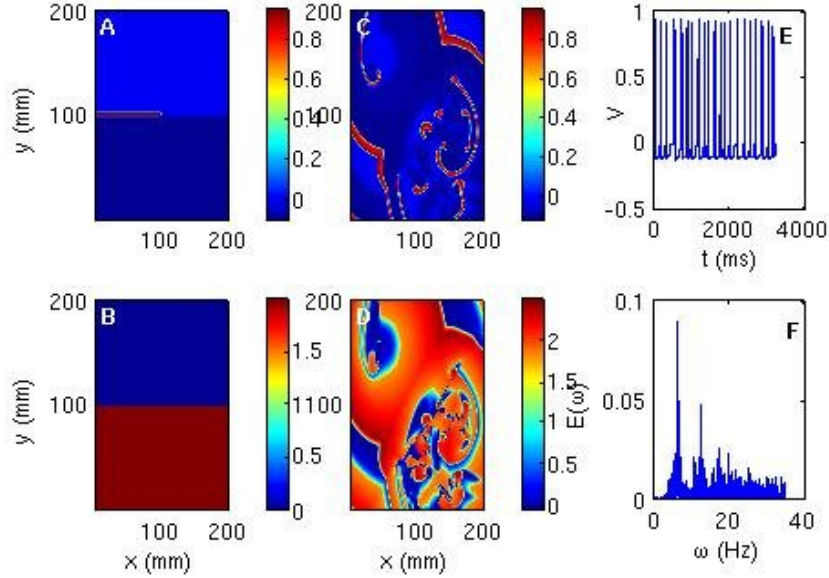

FIG. 2. Spiral-wave break up in the Panfilov Model: Pseudocolour plots of the initial conditions used in our simulation for the fields (A) $V$ and (B) $g$. This initial condition leads to the formation of a spiral wave that breaks down eventually leading to spiral turbulence for (C) $V$ and (D) $g$ fields after $t=2750$ ms. (E) The time series of $V$ from the point (100 mm, 100 mm) and (F) the associated power spectrum [computed after eliminating transients, i.e., the first 50000 time steps (5500 ms) from a time series of 311 424 time steps (34256.64 ms)]; this broad-band power spectrum is characteristic of a chaotic state.

We initialize Eq.(1) at $t=0$ with the following broken wave-front (Fig. 2 A and B): $g=2$ for $0 \leq x \leq L$ and $0 \leq y \leq \frac{L}{2}$, and $g=0$ elsewhere ($L$ is the linear system size); $V=0$ everywhere except for $y=(\frac{L}{2}+1)$ and $0 \leq x \leq \frac{L}{2}$, where $V=0.9$. We observe spiral turbulence if $L > 128$ mm and $\varepsilon_1 = 1/75$; as $\varepsilon_1$ decreases, the spiral pitch decreases and, eventually, broken spirals are obtained; e.g., if $\varepsilon_1 \cong 0.01$, a state containing broken spirals is observed (Figs. 2B and C). However, if $\varepsilon_1 \geq 0.02$, Eq.(1) displays rigidly rotating spirals and, as discussed in Pandit *et al* [6] . Furthermore the breakup of spirals, for $\varepsilon_1 > 1/75$, is associated with the onset of spatiotemporal chaos [6]; spatiotemporally chaotic behaviour is, strictly speaking, a transient, but the duration $\tau_L$ of this transient increases $L$ of the system; for $t > \tau_L$, a quiescent state with $V=g=0$ is obtained. If $\varepsilon_1 > 1/75$ and $L>128$, $\tau_L$ is sufficiently large and we find a nonequilibrium statistical steady with spatiotemporal chaos[6,29], i.e., the number of positive Lyapunov exponents increases with $L$, as does the Kaplan-Yorke dimension $D_{KY}$. Thus VF in the Panfilov model is associated with spiral turbulence and spatiotemporal chaos, initiated by the breakup of spiral waves. Similar spiral turbulence has been observed in more realistic models for ventricular tissue [14, 22] like the Luo-Rudy I model; to the best of our knowledge, Lyapunov spectra and $D_{KY}$ have not been obtained for these models.

The initial condition for transmembrane potential used in LRI models is shown in Fig. 3 A; in the absence of any obstacle in the medium this develops as shown in Figs. 3 B, and C.

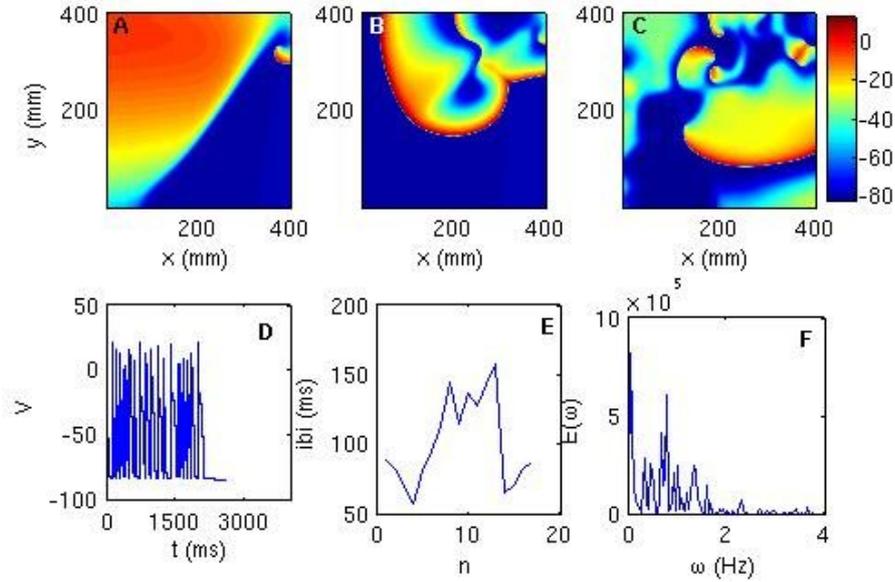

FIG. 3. Representative plots from our simulations of the LRI model for a simulation domain with $L$=90 mm. Pseudocolour plots of the transmembrane voltage $V$: (A) initial condition ($t$=0); (B) after 600 ms; and (C) after 1000 ms. Note that spiral wave has broken in (B) and (C). (D) The local time series of $V$, taken from the point (45 mm, 45 mm); (E) the inter-beat interval (IBI), between two successive spikes in (D), versus the spike number $n$; and (F) the power spectrum associated with the time series of (D). The fluctuations of the IBI and the broad-band power spectrum are signatures of the underlying spiral turbulence.

We model conduction inhomogeneities by setting $D$=0 in regions with obstacles; in all other parts of the simulation domain $D$ is a constant (1 for the Panfilov model and between 0.0005 cm$^2$/s and 0.001 cm$^2$/s for the LRI model). In the studies we present here the obstacle is taken to be a square of side $l$ mm, with 10 mm $\leq l \leq$ 40 mm. For both models we use no-flux (Neumann) boundary conditions on the edges of our simulation domain and on the boundaries of the obstacle (or obstacles).

## Obstacles in the Medium

We have carried out several studies designed to elucidate the effects of conduction inhomogeneities in the Panfilov and LRI models. These are described in detail in Ref. [22]. Here we give an overview of these results and present some new studies that we have carried out on the interplay of two conduction inhomgeneities with spiral waves.

We first examine the dependence of spiral-wave dynamics by fixing the position of the obstacle (cf., Ikeda *et al* [7] for similar experiments) and changing the obstacle size: We place a square obstacle of side $l$ in the Panfilov model in a square simulation domain with side $L$=200 mm. With the bottom-left corner of the obstacle at (50 mm, 100 mm) spiral turbulence (ST) persists if $l \leq (40-\Delta)$ mm; ST gives way to a quiescent state (Q) with no spirals if $l$= 40mm, and eventually to a state with a single rotating spiral (RS) anchored at the obstacle if $l \geq (40+\Delta)$ mm. We have taken $l$ from 2 mm to 80 mm in steps of $\Delta$= 1 mm. Thus we see a clear transition from ST to RS, with these two states separated by a state Q with no spirals.

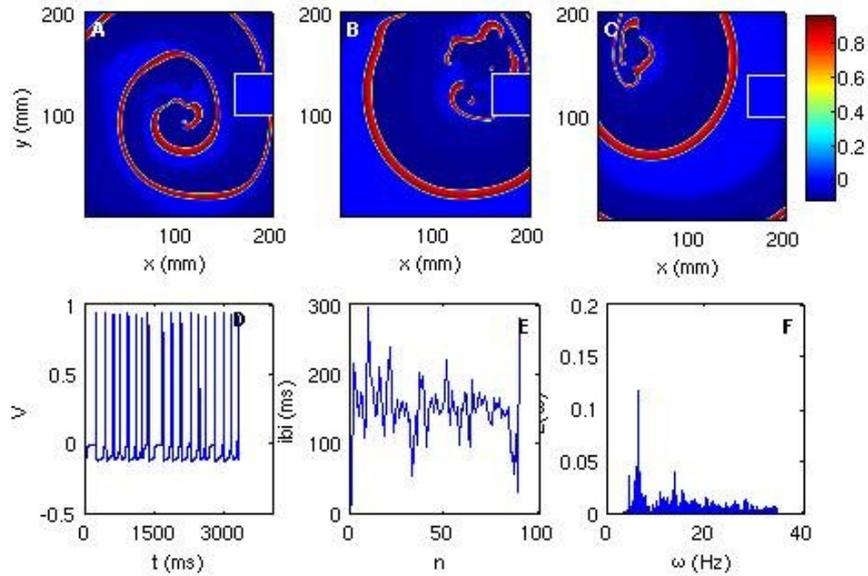

FIG. 4. Spiral turbulence in the Panfilov model with a square obstacle of side 40 mm placed with its lower-left corner at (100 mm, 160 mm). We start with the initial condition of Fig. 2. The spiral wave first gets attached to the obstacle but eventually break up occurs as illustrated in the pseudocolour plots of *V* in (A), (B), and (C) at 1100 ms, 1650 ms, and 2750 ms, respectively. The local time series for *V*, the associated inter-beat interval (IBI), and the power spectrum from a representative point in the simulation domain are shown in (D), (E), and (F), respectively; the nonperiodic behaviour of the IBI and the broad-band nature of the power spectrum are characteristic of the underlying spiral turbulence.

The final state of the system depends on its size and also on where the obstacle is placed with respect to the tip of the initial wavefront. Even a small obstacle, placed near this tip [obstacle with $l$=10 mm at (100 mm, 100 mm)], can prevent the spiral from breaking up; but a bigger obstacle, placed far away from the tip [obstacle with $l= 75$ mm at (125 mm, 50 mm)], does not affect the spiral.

In Ref. [22] we have explored in detail how the position of the obstacle changes the final state; here we give an overview of these results only for the two-dimensional case. One of the remarkable conclusions of our study is that the final state of the system depends sensitively on the position of the obstacle. This is illustrated clearly in Figs. 4-6 for the Panfilov model and Figs. 7- 9 for the LRI model. As these figures show, we can obtain ST, RS, or Q states in both these models depending on where the obstacle is placed; pseudocolour plots of $V$ show broken spirals, one rotating spiral, or no spirals in ST, RS, and Q states, respectively; the time series, inter-beat intervals, and power spectra (from a representative point in the simulation domain) are also shown in Figs. 4-9 for these three states.

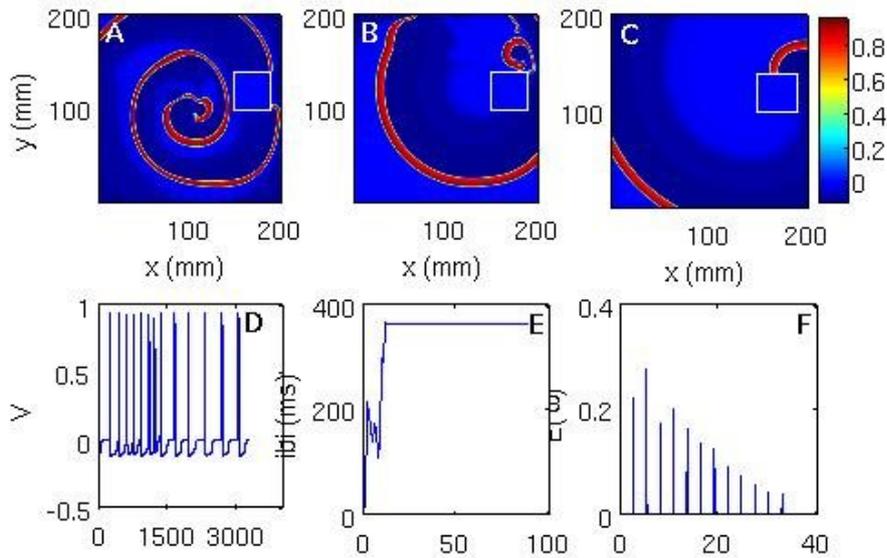

FIG. 5. Pseudocolour plots illustrating the attachment of a spiral wave to a square obstacle of side 40 mm, in the Panfilov model, with its lower-left corner is at (100 mm, 150 mm) and the initial condition of Fig. 2. (A), (B), and (C) show pseudocolour plots of $V$ at 1100 ms, 1650 ms, and 2750 ms, respectively. The wave gets attached to the obstacle after about 9 spiral rotations ($\cong 1800$ ms). The resulting periodic behviour is reflected in (D) the time series of $V$ and (E) the plot of the inter-beat interval IBI versus the spike number $n$: after transients the IBI settles to a constant value of 363 ms; (F) the power spectrum has discrete peaks at the fundamental frequency $\omega_f = 2.74$ Hz and its harmonics.

Our conclusions are summarized in the stability diagram of Fig. 10. This gives the effect of a square conduction inhomogeneity, with side $l=40$ mm, in a square simulation domain with side $L= 200$ mm. The colours of the small squares (of side $l_p$) in Fig. 10 indicate the final state of the system when the position of the bottom-left corner of the obstacle coincides with that of the small square; red, blue, and green denote, respectively, spiral turbulence (ST), a single rotating spiral (RS) anchored at the obstacle, and a quiescent state (Q). The boundaries between ST, RS, and Q appear to be fractal-like as suggested by the enlarged versions of Fig. 10A in Figs. 10B and C.

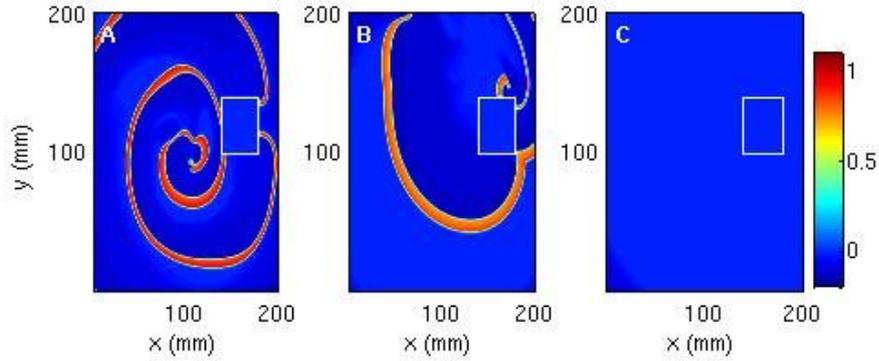

FIG. 6. Pseudocolour plots illustrating how, in the Panfilov model, a spiral wave moves away from the simulation domain if a square obstacle of side 40 mm is placed with its lower-left corner at (100 mm, 140 mm) and we use the initial condition of Fig. 2. (A), (B), and (C) shows pseudocolour plots of $V$ at 1100 ms, 1650 ms and 2750 ms, respectively.

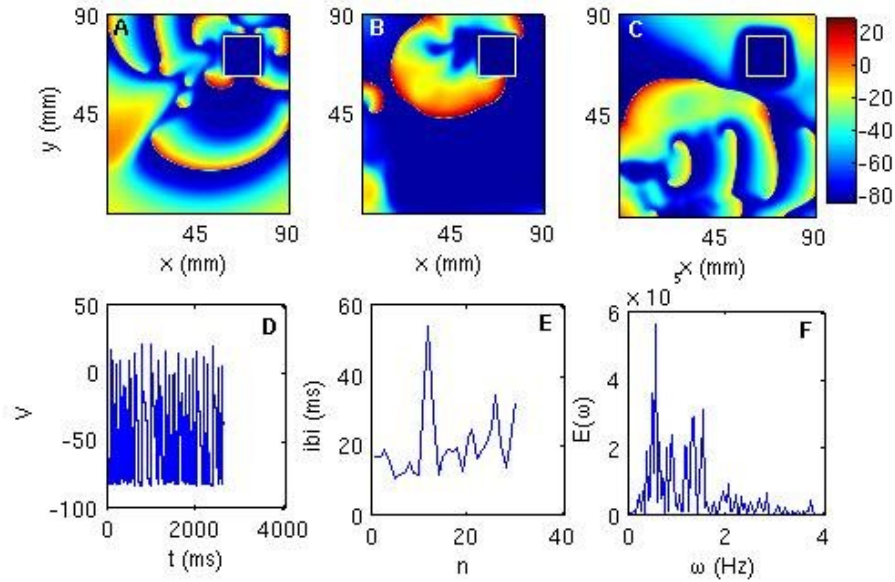

FIG. 7. Spiral turbulence in the presence of an obstacle in the LRI model: We illustrates how the initial condition (Fig. 3A) evolves in the presence of an obstacle of side $l$=18 mm placed with its bottom-left corner at (58.5 mm, 63 mm); spiral turbulence persists in this case; (A), (B), and (C) shows pseudocolour plots of $V$ at 200, 600, and 1000 ms, respectively. (D) The local time series from (45 mm, 45

mm); (E) the inter-beat interval (IBI) for this time series; and (F) the power spectrum for the time series in (D) [from 261424 timesteps (i.e., 2614.24 ms)].

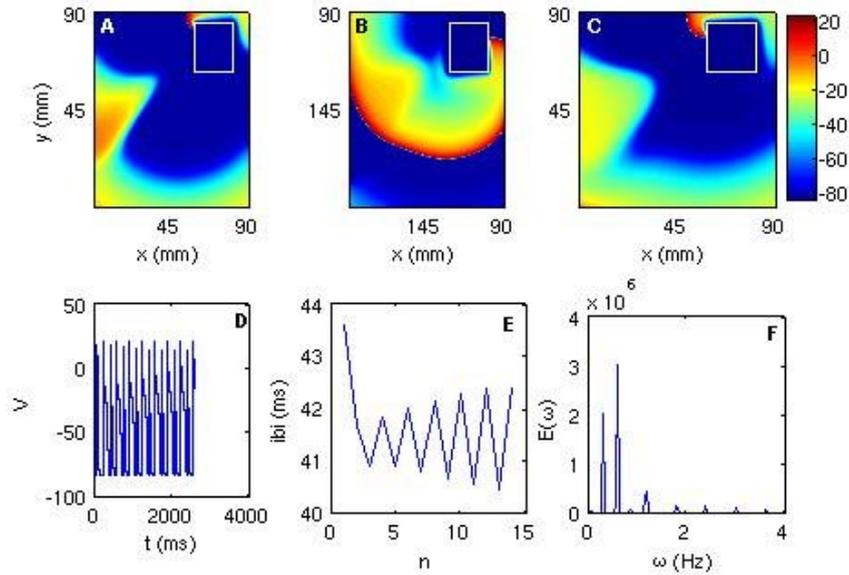

FIG. 8. Spiral-wave anchoring in the presence of an obstacle in the LRI model: We illustrate how the initial condition (Fig. 3A) evolves in the presence of an obstacle of side $l$=18mm with its bottom-left corner at (58.5 mm, 63 mm) to a spiral wave anchored to the obstacle. (A), (B), and (C) show pseudocolour plots of $V$ at 200, 600, and 1000 ms, respectively. (D) The local time series from (45 mm, 45 mm); (E) the inter-beat interval (IBI) versus the spike number $n$ for this time series; and (F) the power spectrum for the time series in (D) [from 261424 timesteps (i.e., 2614.24 ms)].

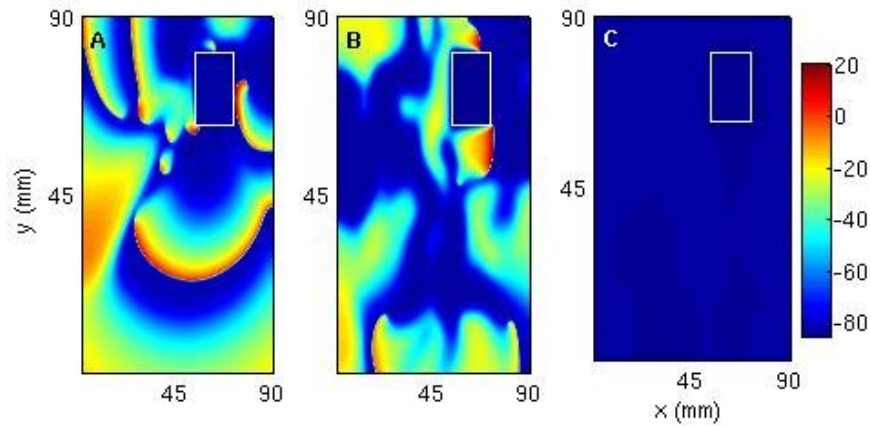

FIG. 9. A spiral-wave moving away from the simulation domain in the presence of an obstacle in the LRI model: We illustrate how the initial condition (Fig. 3A) evolves in the presence of an obstacle of side $l$=18mm with its bottom-left corner at (54 mm,63 mm). (A), (B), and (C) show pseudocolour plots of $V$ after 200 ms, 600 ms, and 1000 ms, respectively.

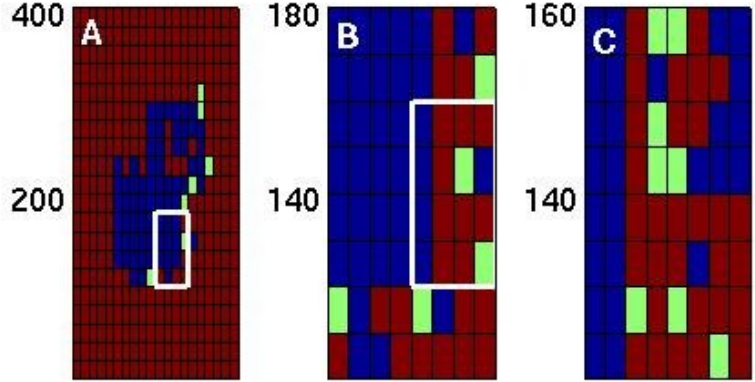

FIG. 10. Panfilov-model stability diagram: The effect of a square conduction inhomogeneity with side $l$=40 mm in a square simulation domain with side $L$= 200 mm. In this diagram the colours of the small squares (of side $l_p$) indicate the final state of the system when the position of the bottom-left corner of the obstacle coincides with that of the small square; the colours red, blue, and green denote, respectively, spiral turbulence (ST), a single rotating spiral (RS) anchored at the obstacle, and a quiescent state (Q). (A) for $l_p$ =10 mm. We expose the fractal-like structure of the boundaries between ST, RS, and Q by zooming in on small subdomains encompassing parts of these interfaces (white boundaries in (A) and (B)) with (B) $l_p$=5 mm, (C) and $l_p$ =2.5 mm.

The next issue we address is the dependence of spiral-wave dynamics in these models on additional obstacles change. We place the first obstacle A at $(x_A, y_A)$ and the next obstacle B at $(x_B, y_B)$. For simplicity we show results only from our studies of the Panfilov model in two dimensions. The effects of obstacle A are summarized in the stability diagram of Figure 11 A that uses the same colour code as Fig. 10; we consider only 18 positions for the obstacle A. The obstacle B is now kept at (120 mm, 110 mm). If only obstacle B were in this position without obstacle A, spiral break up would have continued. Now we study the combined effects of both these obstacles on the spiral waves.

We note first that, in most cases, effects of one of the obstacles dominate spiral-wave dynamics. For example, when obstacle A alone is placed at $(x_A =120$ mm, $y_A =60$ mm) the spiral wave is anchored to it; the addition of a second obstacle does not change this; similarly, when obstacle A alone is at $(x_A = 90$ mm, $y_A =50)$, the spiral moves away from the simulation domain; this too is not affected by the presence of the second obstacle. However, if obstacle A is at $(x_A =110$ mm, $y_A =50$ mm), the spiral wave gets anchored to it; now if the second obstacle B is placed at $(x_B =120$ mm, $y_B$=110 mm), the spiral waves breaks up. The most interesting case occurs if obstacle A is at (120 mm, 40 mm) and obstacle B at (120 mm, 110 mm); in the absence of the other, each one of these obstacles causes the spiral waves to break up; but together they cause these waves to move away from the medium!

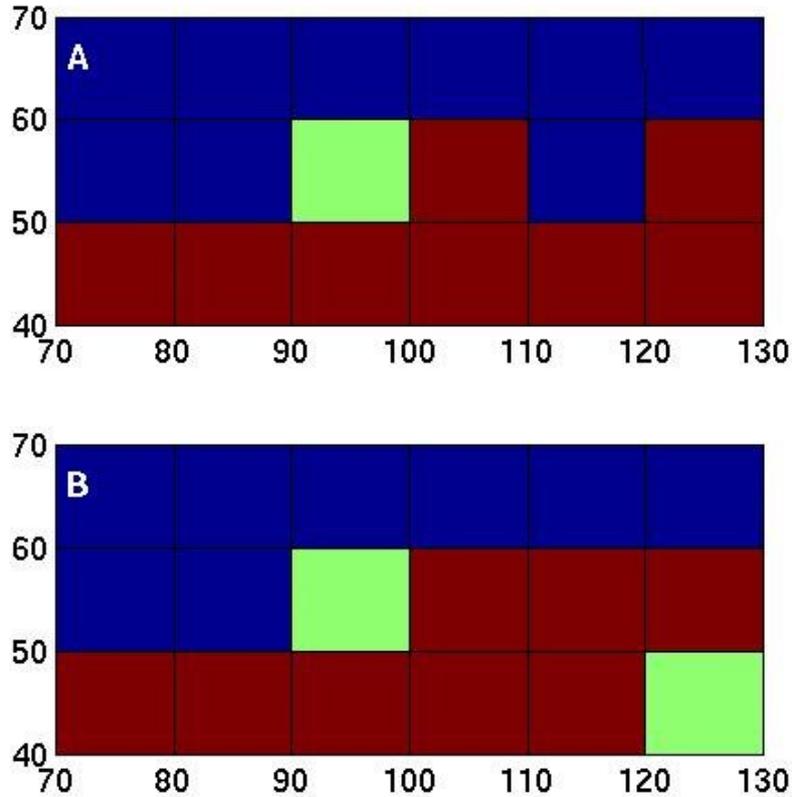

FIG. 11. (A) Stability diagram for the Panfilov model ($L=200$ mm and $l_p = 10$ mm) with a square obstacle A, of side $l=40$ mm. The colour code is the same as the one used in Fig.10. (B) The analogue of (A) but with two square obstacles A and B of the same size. Obstacle B is placed at (120 mm, 110 mm). The colour of a square shows the final state of the system when the left bottom corner of obstacle A coincides with that of the square.

## Conclusions

We have given an overview of studies of the effects of conduction inhomogeneities on spiral-wave dynamics in models for ventricular tissue. In particular, we have presented our extensive and systematic work on this problem in the Panflov and LRI models. The most remarkable qualitative conclusion of our study is that spiral-wave dynamics in these models depends very sensitively on the positions of conduction inhomogeneities. All possible behaviours, ST, RS, and Q are possible as shown in the stability diagrams of Figs. 9 and 10; and the boundaries between these appear to be fractal like. As we have argued in Ref. [22], this must arise because of an underlying fractal basin boundary between the domains of attraction of the ST, RS, and Q states in our high-dimensional dynamical systems, the Panfilov and LRI models (strictly speaking these are infinite-dimensional dynamical systems since they are PDEs). Normally one goes from one domain of attraction to another in a dynamical system by changing the initial condition. Here we do not change the initial condition but we modify the dynamical system slightly by changing the position of the obstacle. Our studies show that the final state of the system depends very sensitively on this change.

Our study has obvious and important implications for cardiac arrhythmias. It suggests that such arrhythmias must depend sensitively on the positions of conduction inhomogeneities in cardiac tissue. We believe this qualitative insight will help in the development of low-amplitude defibrillation schemes that can work even in the presence of inhomogeneities in cardiac tissue.

**Acknowledgements**

We thank DST, UGC, and CSIR (India) for support, and SERC (IISc) for computational resources.

## Appendix : The Luo Rudy I Model

In the Luo-Rudy I (LR I) model there are six components of the ionic current, which are formulated mathematically in terms of Hodgkin-Huxley-type equations. The partial differential equation for the transmembrane potential *V* is

$$\frac{\partial V}{dt} = \frac{I_{LR}}{C} = D\nabla^2 V. \qquad (2)$$

Here $I_{LR}$ is the instantaneous, total ionic-current density. The subscript *LR* denotes that we use the formulation of the total ionic current described by the Luo-Rudy Phase I (LR1)model[25]. In the LR1 model, $I_{LR} = I_{Na} + I_{si} + I_K + I_{K1} + I_{Kp} + I_b$, with the current densities $I_{Na}$ (fast inward Na$^+$), $I_{si}$ (slow inward), $I_K$ (slow outward time-dependent K$^+$), $I_{K1}$ (time-independent K$^+$), $I_{Kp}$ (plateau K$^+$), $I_b$ (total background), given by:

$$I_{Na} = G_{Na}m^2hj(V - E_{Na})$$
$$I_{si} = G_{si}df(V - E_{si})$$
$$I_K = G_k xx_i(V - E_K)$$
$$I_{K_1} = G_K K_{1\infty}(V - E_{K_1})$$
$$I_{Kp} = G_{Kp}K_p(V - E_{Kp})$$
$$I_b = 0.03921(V + 59.87)$$

and $K_{1\infty}$ is the steady-state value of the gating variable $K_1$. All current densities are in units of μA/cm$^2$, voltages are in mV, and $G_\xi$ and $E_\xi$ are, respectively, the ion-channel conductance and reversal potential for the channel ξ. The ionic currents are determined by the time-dependent ion-channel gating variables *h, j, m, d, f, x, $x_i$, $K_p$* and *$K_1$* generically denoted by ξ, which follow ordinary differential equations of the type

$$\frac{d\xi}{dt} = \frac{\xi_\infty - \xi}{\tau_\xi}$$

where $\xi_\infty = {\alpha_\xi}/{(\alpha_\xi + \beta_\xi)}$ is the steady-state value of $\xi$ and $\tau_\xi = \frac{1}{\alpha_\xi + \beta_\xi}$ is its time constant. The voltage-dependent rate constants, $\alpha_\xi$ and $\beta_\xi$, are given by the following empirical equations:

$\alpha_h = 0$, if $V \geq -40mV$,

$\quad = 0.135 \exp\left[-0.147(V+80)\right]$, Otherwise;

$\beta_h = \dfrac{1}{0.13\left(1+\exp\left[-0.09(V+10.66)\right]\right)}$, if $V \geq -40mV$

$\quad = 3.56\exp[0.079V] + 3.1\times10^{-5}\exp[0.35V]$, otherwise;

$\alpha_j = 0$, if $V \geq -40mV$

$\quad = \left[\dfrac{\left(\exp[0.2444V] + 2.732\times10^{-10}\exp[-0.04391V]\right)}{-7.865\times10^{-6}\left\{1+\exp\left[0.311(V+79.23)\right]\right\}}\right](V+37.78)$,

otherwise;

$\beta_j = \dfrac{0.3\exp\left[-2.535\times10^{-7}V\right]}{1+\exp\left[-0.1(V+32)\right]}$, if $V \geq -40mV$

$\quad = \dfrac{0.1212\exp[-0.01052V]}{1+\exp\left[-0.1378(V+40.14)\right]}$, otherwise;

$$\alpha_m = \frac{0.32(V+47.13)}{1-\exp[-0.1(V+47.13)]};$$

$$\beta_m = 0.08\exp[-0.0909V];$$

$$\alpha_d = \frac{0.095\exp[-0.01(V-5)]}{1+\exp[-0.072(V-5)]};$$

$$\beta_d = \frac{0.07\exp[-0.017(V+44)]}{1+\exp[0.05(V+44)]};$$

$$\alpha_f = \frac{0.012\exp[-0.008(V+28)]}{1+\exp[0.15(V+28)]};$$

$$\beta_f = \frac{0.0065\exp[-0.02(V+30)]}{1+\exp[-0.2(V+30)]};$$

$$\alpha_x = \frac{0.0005\exp[0.083(V+50)]}{1+\exp[0.057(V+50)]};$$

$$\beta_x = \frac{0.0013\exp[-0.06(V+20)]}{1+\exp[-0.04(V+20)]};$$

$$\alpha_{K1} = \frac{1.02}{1+\exp[0.2385(V-E_{K1}-59.215)]};$$

$$\beta_{K1} = \left[\frac{0.49124\exp[0.08032(V-E_{K1}+5.476)]}{1+\exp[-0.5143(V-E_{K1}+4.753)]} + \exp[0.06175(V-E_{K1}-594.31)]\right].$$

The gating variables $x_i$ and $K_p$ are given by

$$x_i = \frac{2.837\exp 0.04(V+77)-1}{(V+77)\exp 0.04(V+35)}, \text{ if } V \succ -100 mV,$$

$$= 1, \text{ otherwise};$$

$$K_p = \frac{1}{1+\exp[0.1672(7.488-V)]}.$$

The values of the channel conductances $G_{Na}$, $G_{si}$, $G_K$, $G_{K1}$, and $G_{Kp}$ are 23, 0.07, 0.705, 0.6047 and 0.0183 mS/cm$^2$, respectively. The reversal potentials are $E_{Na}$=54.4 mV, $E_K$=-77 mV, $E_{K1}$=$E_{Kp}$=-87.26 mV, $E_b$=-59.87 mV, and $E_{si}$=7.7-13.0287lnCa, where $Ca$ is the calcium ionic concentration satisfying

$$\frac{dCa}{dt} = -10^{-4} I_{si} + 0.07 \left(10^{-4} - Ca\right).$$

The times $t$ and $\tau_\xi$ are in ms; the rate constants $\alpha_\xi$ and $\beta_\xi$ are in ms$^{-1}$.

## References

[1] J. R. Hampton, *The E C G Made Easy*, Churchill Livingstone (2003)

[2] M. C. Cross, and P. C. Hohenberg, *Pattern formation outside equilibrium*, Rev. Mod. Phys., **65**, 851 (1993)

[3] A. T. Winfree, Evolving perspectives during 12 years of electrical turbulence, Chaos, **8**,1 (1998).

[4] A. T. Winfree, *When Time Breaks Down,* Princeton University Press (1987)

[5] J. Jalife, R. A. Gray, G. E. Morley, and J. M. Davidenko, *Spiral breakup as a model of ventricular fibrillation*, Chaos, **8**, 57 (1998).

[6] R. Pandit, A. Pande, S. Sinha, and A. Sen, *Spiral turbulence and spatiotemporal chaos: characterization and control in two excitable media,* Physica A **306**, 211 (2002).

[7] T. Ikeda, M. Yashima, T. Uchida, D. Hough, M. C. Fishbein, W. J. Mandel, P.-S. Chen, and H. S. Karagueuzian, *Attachment of meandering reentrant wave fronts to anatomic obstacles in the atrium: role of obstacle size*, Circ. Res. **81**, 753 (1997).

[8] J. M. Davidenko, A. V. Pertsov, R. Salomonsz, W. T. Baxter, and J. Jalife, *Stationary and drifting spiral waves of excitation in isolated cardiac muscle*, Nature, **355**, 349 (1929).

[9] M. Valderrabano, P. S. Chen, and S. F. Lin, *Spatial distribution of phase singularities in ventricular fibrillation,* Circulation, **108**, 354 (2003).

[10] M. Valderrabano, M.-H. Lee, T. Ohara, A. C. Lai, M. C Fishbein, S.-F. Lin, H. S. Karagueuzian, and P. S. Chen, *Dynamics of intramural and transmural reentry during ventricular fibrillation in isolated swine ventricles,* Circ. Res. **88**, 839-848 (2001).

[11] T.-J. Wu, J. J. Ong, C. Hwang, J. J. Lee, M. C. Fishbein, L. Czer, A. Trento, C. Blanche, R. M. Kass, W. J. Mandel, *et al. Characteristics of wave fronts during ventricular fibrillation in human hearts with dilatedcardiomyopathy: role of increased fibrosis in the generation of reentry,* J. of Am. Coll. of Card., **32**, 187 (1998).

[12] T. Ohara, K. Ohara, J.-M. Cao, M.-H. Lee, M. C. Fishbein, W. J. Mandel, P.-S. Chen, and H. S. Karagueuzian, *Increased wave break during ventricular fibrillation in the epicardial border zone of hearts with healed myocardial infarction,* Circulation.**103**, 1465-1472 (2001).

[13] S.-M. Hwang, T. Y. Kim, and K. J. Lee, *Complex-periodic spiral waves in confluent cardiac cell cultures induced by localized inhomogeneities,* Proc. Natl. Acad. Sci. USA, **102**,10363 (2005).

[14] F. Xie, F. Qu and A. Garfinkel, *Dynamics of reentry around a circular obstacle in cardiac tissue*, Phys. Rev. E, **58**, 6355 (1998)

[15] A. M. Pertsov, J. M. Davidenko, R. Salomonsz, W. T. Baxter and J. Jalife, *Spiral waves of excitation underie reentrant activityisolated cardiac muscle*, Circ. Res. **72**, 631 (1993)

[16] M. Valderrabano, Y. –H. Kim, M. Yasima, T. –J. Wu, H. S. Karagueuzian and P. –S. Chen, *Obstacle-induced transition from ventricular fibrillation to tachycardia in isolated swine right ventricles*, Journ. Of Am. Coll. Cardiol. **36**, 2000 (2000)

[17] Y. H. Kim, F. Xie, M. Yashima, T.-J. Wu, M. Valderrabano, M.-H. Lee, T. Ohara, O. Voroshilovsky, R. N. Doshi, M. C. Fishbein, *et al. Role of papillary muscle in the generation and maintenance of reentry during ventricular tachycardia and fibrillation in isolated swine right ventricle,* Circulation. **100**, 1450 (1999).

[18] J. M. Starobin, and C. F. Starmer, *Boundary-layer analysis of waves propagating in an excitable medium: medium conditions for wave-front-obstacle separation,* Phys. Rev. E, **54**, 430 (1996).

[19] S. Takagi, A. Pumir, D. Pazo, I. Efimov, V. Nikolski, and V. Krinsky, *Unpinning and removal of a rotating wave in cardiac tissue,* Phys. Rev. Lett. **93**, 058101 (2004).

[20] A. V. Panfilov, *Spiral breakup in an array of coupled cells: the role of the intracellular conductance,* Phys. Rev. Lett. **88**, 118101 (2002).

[21] K. H. W. J. ten Tusscher, and A.V. Panfilov, *Influence of nonexcitable cells on spiral breakup in two-dimensional and three-dimensional excitable media,* Phys. Rev. E **68**, 062902 (2003).

[22] T. K. Shajahan, S. Sinha, and R. Pandit, *Spiral-wave dynamics depend sensitively on inhomogeneities in mathematical models of ventricular tissue*, Phys. Rev. E, **75,** 011929 (2007).

[23] A. V. Panfilov, *Spiral breakup as a model of ventricular fibrillation*, Chaos **8**, 57 (1998).

[24] A. V. Panfilov and P. Hogeweg, Phys. Lett. A, **176**, 295 (1993).

[30] C. H. Luo and Y. Rudy, *A model of the ventricular cardiac action potential: depolarization, repolarization, and their interaction,* Circ. Res. **68**, 1501 (1991).

[26] T. K. Shajahan, S. Sinha, and R. Pandit, *Ventricular fibrillation in a simple excitable medium model of cardiac tissue,* Int. Journ. of Modern Physics B, **17**, 5645 (2003).

[27] T. K. Shajahan, S. Sinha, and R. Pandit, *Spatiotemporal chaos and spiral turbulence in models of cardiac arrhythmias,* Proc. Indian Natl. Sci. Acad., **71** A, 4757 (2005).

[28] A. Pande, and R. Pandit, *Spatiotemporal chaos and nonequilibrim transitions in a model excitable medium*, Phys. Rev. E, **61,** 6448 (2000).

[29] S. Sinha, A. Pande, and R. Pandit, *Defibrillation via the elimination of spiral turbulence in a model for ventricular fibrillation,* Phys. Rev. Lett. **86**, 3678 (2001).